\pdfoutput=1

\documentclass[a4,11pt]{article}

\usepackage{graphicx,sidecap}
\usepackage{xcolor}
\usepackage{lmodern}
\usepackage{epsfig,epsf,subfigure,url}
\usepackage{natbib}
\usepackage{bm}
\usepackage{times}
\usepackage{pstricks-add}

\RequirePackage{amsthm,amsmath,amsfonts,amssymb}
\RequirePackage[colorlinks,citecolor=blue,urlcolor=blue]{hyperref}
\RequirePackage{graphicx}

\usepackage{enumerate,mathrsfs,comment}
\usepackage{bbm}
\usepackage{enumitem}
\usepackage[OT1]{fontenc}

\usepackage[protrusion=false,expansion=true]{microtype}
\usepackage{mathtools}

\RequirePackage{amsmath}
\RequirePackage{amssymb}
\RequirePackage{amsthm}
\RequirePackage{bm}
\usepackage{natbib}
\usepackage{bm}
\usepackage{multirow}
\usepackage{graphicx}
\usepackage{subfigure}
\usepackage{makecell}
\usepackage{booktabs}
\usepackage{array}
\usepackage{algorithm}
\usepackage{algorithmic}
\usepackage{dsfont}

\usepackage{amsgen,amsmath,amstext,amsbsy,amsopn,amssymb,amscd}%
\usepackage{color}

\newtheorem{theorem}{Theorem}[section]

\newtheorem{lemma}{Lemma}[section]

\newtheorem{definition}{Definition}[section]

\newcommand{\fdr}{\textsc{fdr }}

\newcommand{\fdrend}{\textsc{fdr}}
\newcommand{\fnrend}{\textsc{fnr}}

\newcommand{\mfdr}{m\textsc{fdr }}
\newcommand{\mfnr}{m\textsc{fnr }}
\newcommand{\mfdrend}{m\textsc{fdr}}
\newcommand{\mfnrend}{m\textsc{fnr}}
\newcommand{\atpend}{\textsc{atp}}
\newcommand{\afpend}{\textsc{afp}}
\newcommand{\atp}{\textsc{atp }}
\newcommand{\afp}{\textsc{afp }}

\newcommand{\iid}{\stackrel{\mathrm{iid}}{\sim}}

\newcommand{\vI}{\mathbf{I}}

\newcommand{\vx}{\boldsymbol{x}}
\newcommand{\vX}{\boldsymbol{X}}
\newcommand{\vy}{\boldsymbol{y}}
\newcommand{\vY}{\boldsymbol{Y}}

\newcommand{\vzero}{\boldsymbol{0}}

\newcommand{\bbeta}{\boldsymbol{\beta}}
\newcommand{\bdelta}{\boldsymbol{\delta}}

\newcommand{\btheta}{\boldsymbol{\theta}}

\newcommand{\bepsilon}{\boldsymbol{\epsilon}}

\newcommand{\eps}{\epsilon}

\newcommand{\lambdapp}{ \log(\frac{\lambda\pi_0}{\pi_1})\frac{1}{\tau_p}+\frac{\tau_p}{2} }
\newcommand{\lambdapn}{ \log(\frac{\lambda\pi_0}{\pi_1})\frac{1}{\tau_p}-\frac{\tau_p}{2} }

\newcommand{\optrate}{ \frac{r^2+(\vartheta-\zeta)^2+2r(\vartheta+\zeta)}{4r} }
\newcommand{\cf}{\bar{F}}

\title{Optimal Multiple Testing in High-Dimensional Regression
}

\author{
Pengsheng Ji \thanks{Associate Professor, Department of Statistics, University of Georgia}, $\quad$
    Zhigen Zhao\thanks{Associate Professor, Department of Statistics, Operations, and Data Science, Temple University}, $\quad$
}
\date{}

\begin{document}
\maketitle
\noindent {\bf Abstract:} In the high dimensional regression analysis when the number of predictors is much larger than the sample size, an important question is to select the important variable which are relevant to the response variable of interest. Variable selection and the multiple testing are both tools to address this issue. However, there is little discussion on the connection of these two areas. When the signal strength is strong enough such that the selection consistency is achievable, it seems to be unnecessary to control the false discovery rate. In this paper, we consider the regime where the signals are both rare and weak such that the selection consistency is not achievable and propose a method which controls the false discovery rate asymptotically. It is theoretically shown that the false non-discovery rate of the proposed method converges to zero at the optimal rate. Numerical results are provided to demonstrate the advantage of the proposed method.


\section{Introduction}\label{sec:intro}
High-dimensional data analysis has become an increasingly active area of research in analyzing data from many modern scientific research areas.
In this paper, we consider a setting in which there is one continuous response variable $Y$ and $p$ predictors for each subject, out of total $n$ subjects (with $p$ being much larger than $n$).
We consider the following regression model
\begin{equation}\label{model:reg}
\vY=\vX\bbeta+\bepsilon,
\end{equation}
where $\vY=(Y_1, Y_2,\cdots, Y_n)^T$, $\vX=(x_{ij})$, and $\bepsilon\sim N(\vzero,\sigma^2\vI_n)$. In this formula, $i=1,2,\cdots, n$ and $j=1,2,\cdots, p$ with $x_{ij}$ being the value corresponding to the $j$-th predictor of the $i$-th subject.
Let $\vX_{i\cdot}$ (the $i$-th row of $\vX$) be the values for the $i$-th subject; and $\vX_{\cdot j}$, the $j$-th column of $\vX$, be the values corresponding to the $j$-th predictor for all the subjects.
In many applications, it is known that the vector $\bbeta$ is sparse in the sense that the majority of the coordinates are zero.
One objective is to identify as many non-zeros as possible, subject to control of the false positives.
Namely, for each predictor $\vX_{\cdot j}$, we want to test whether this predictor has a non-zero effect on the response $\vY$. This can be described by the following hypotheses:
\[
H_j: \beta_j=0, \ j=1,2,\cdots,p.
\]
Let $\btheta=(\theta_1,\cdots,\theta_p)$ where $\theta_j=\mathbbm{1}(\beta_j\neq 0)$ and $\bdelta=(\delta_1,\cdots,\delta_p)\in\{0,1\}^p$ be a decision based on the data where $\delta_j=1$ if one decides to reject $H_j$ and $0$ if not.
Ideally, one would like to reject all those hypotheses with $\theta_j =1 $ while accepting all the others with $\theta_j=0$ with high probability, i.e.
\begin{align}
\lim_{p\to\infty} P\left( \sum_j \mathbbm{1}(\theta_j\neq \delta_j) =0 \right)\to 0.
\label{eq:oracleProperty}
\end{align}
This goal is known as the ``selection consistency'' or ``oracle property'' in the variable selection literature, and requires
the signals are sufficiently strong. See, e.g., \cite{Fan:Li:2001}, \cite{Zhao:Yu:2006}, \cite{Meinshausen:Buhlmann:2006}, \cite{Meinshausen:Buhlmann:2010}, and \cite{Zou:2006}.

However, in many modern applications, a large $p$ usually means that signals are rare, and a small $n$ usually means that the signals are weak.
Under the regime of rare and weak signals, the conditions required for the selection consistency are, unfortunately, too strong to be true \citep{Ji:Jin:2012, Zhang:Zhang:2014}.
Therefore, it is scientifically more relevant to allow a certain number of false positives as long as a chosen type I error rate can be controlled at a given level.
Such error rates include, but are not limited to, false discovery rate (\fdrend, \cite{Benjamini:Hochberg:1995}, \cite{Benjamini:Yekutieli:2001}), marginal false discovery rate (\mfdrend, \cite{Genovese:Wasserman:2002}, \cite{Genovese:Wasserman:2004}), and many others. 

In the last decade, many methods have been introduced to control the false discovery rate under the regression model. In \cite{barber2015controlling}, they introduced the knockoff filter which offers non-asymptotic FDR control when $p<n$. There are many further extension of knockoff afterwards. To name a few here, in \citep{barber2019knockoff}, the authors have been extended to high-dimensional setting. In \cite{candes2018panning}, the authors considered the model-X knockoff filter. Many other extensions including but not limited to \citep{wang2020power, barber2020robust}. Recently, \cite{xing2021controlling} introduced the Gaussian mirror for FDR control under the linear regression model. It has been extended to the generalized linear regression model \citep{dai2022false}. \cite{Zhao:Xing:2022} has further introduced a model-free FDR control method in the area of sufficient dimension reduction. 

In all the aforementioned work, the main focus is on the validity of the method. There is no theoretical investigation on the optimality. There are some existing optimality results which unfortunately, can not be applied to the regression settings (\cite{Sarkar:2004, Genovese:Wasserman:2002, Genovese:Wasserman:2004, finner2009false}). In this paper, we tackle this issue by borrowing the idea from the optimal variable selection. 

In \cite{Ji:Jin:2012}, the authors initiated the study of the optimality of variable selection using the Hamming error when the ``selection consistency'' is not possible. They proposed the Univariate Penalization Screening (UPS) method and further showed that the UPS procedure achieves the optimal rate under certain conditions. The UPS procedure allows false discoveries as long as the total number of false discoveries is comparable to the optimal risk. However, because the type I and type II errors are treated equally, using this procedure naively for multiple testing makes the \mfdr go to zero, but also substantially limits the power to discover signals.


In this article, we consider the following loss function
\begin{equation}\label{eqn:loss}
  L(\btheta, \bdelta) = \sum_{j=1}^p \lambda (1-\theta_j)\delta_j + \theta_j(1-\delta_j).
\end{equation}
The loss function is a weighted Hamming error which treat the type I and type II errors unequally. For a given weight, an optimal variable selection method which minimizes the weighted Hamming error leads to an optimal multiple testing method with the FDR level relies on this weight though the relation is not tractable. We further develop a method to estimate this weight using the data such that the false discovery rate is controlled at any designated level asymptotically. It is further shown that the false non-discovery rate of the proposed method converges to zero at the optimal rate.

The remaining sections are organized as follows. In Section \ref{subsec:lower}, the rare/weak model is specified and the lower bound of the convergence rate of \mfnr is given. The UPT method is proposed in Section \ref{sec:upper} and the optimality results are provided. Some numerical studies in Section \ref{sec:simulation} demonstrate the performance of the UPT method. The proofs are left in Section \ref{sec:appendix}. For technical convenience, we adopt some similar notations from \cite{Ji:Jin:2012} if needed.


\section{The Rare/Weak Model and Lower Bound of m{\textsc{FNR}}}\label{subsec:lower}

We assume the following regression model with rare and weak signals:
\begin{equation}\label{eqn:model:2}
\left\{ \begin{array}{l}
    \vY=\vX\bbeta + \bepsilon, \epsilon\sim N(0, \vI_n) ;\\
    \beta_j \iid \pi_0 h_0 + \pi_1 h_1, \\
    \textrm{where $h_0$ has the point mass at 0 and $support(h_1)=[-\tau_p, 0)\cup (0, \tau_p]$};\\
    \textrm{$\tau_p=\sqrt{2r\log{p}}$};\\
    \pi_1= p^{-\vartheta}.
\end{array}\right.
\end{equation}
Note that the support of the signal distribution $h_1$ has the order of $\sqrt{\log{p}}$, representing rare and weak signals;
the proportion of signals $\pi_1$ goes to zero as $p$ goes to infinity, indicating rare signals.
This model goes back to \cite{Donoho:Jin:2004} and has been used by \cite{Jager:Wellner:2007}, \cite{Candes:Plan:2009}, \cite{Ingster:Pouet:Tsybakov:2009} and many others.

In this section, we study the lower bound of the rate of the \mfnr of any testing procedure which has \mfdr being controlled at $\alpha$ level.
Assume the loss function (\ref{eqn:loss}),
then the oracle decision is given as
\begin{equation}\label{eqn:optimal:rule}
  \delta_j^{opt} = \mathbbm{1}\left( fdr_j(\vY)\le\frac{1}{1+\lambda} \right),
\end{equation}
where
\begin{equation}\label{def:loc:fdr}
fdr_j(\vY)= P(\theta_j=0|\vY)
\end{equation}
is the generalized local fdr \citep{Efron:2008, Efron:2010b, He:Sarkar:Zhao:2015, Liu:Sarkar:Zhao:2016} for the $j$-th hypothesis, or local index of significance \citep{Sun:Cai:2009}. This consideration is also closely related to the mulitple testing under the empirical Bayes framework \citep{dickhaus2014simultaneous, Zhao:Sarkar:2015, kwon2018f}.

Next, we study the risk of the oracle decision rule (\ref{eqn:optimal:rule}).
\begin{theorem}\label{thm:risk}
Assume Model (\ref{eqn:model:2}) and the loss function (\ref{eqn:loss}). Then the risk of the oracle decision rule (\ref{eqn:optimal:rule}) is given as
\begin{equation*}
 \sum_j E\left\{ \mathbbm{1}\left( fdr_j(\vY)\le \frac{1}{1+\lambda}\right) \left[ (\lambda+1)fdr_j(\vY)   - 1 \right] +\pi_1\right \}.
\end{equation*}
\end{theorem}
Theorem \ref{thm:risk} is a general theorem that can actually be applied to much broader settings than Model (\ref{eqn:model:2}).
Before stating the next theorem about the rate of the risk, we will recall the definition of ``multi-log'' term introduced in \cite{Ji:Jin:2012}.

\begin{definition}
The term $L_p>0$ is a multi-log($p$) term if for any constant $\varkappa>0$,
\[
\lim\limits_{p\to\infty}   L_p p^{\varkappa}=\infty,  \lim\limits_{p\to\infty}  L_p p^{-\varkappa}=0,
\]
\end{definition}


\begin{theorem}\label{thm:fdr:fnr}
Assume Model (\ref{eqn:model:2}) with the following conditions
\begin{align}
\label{eqn:LB:unit:diag}
 \vX'\vX \ \mbox{ has unit diagonals, }
\end{align}
and
\begin{align}
\label{eqn:LB:signal}
  h_1 \ \mbox{ is supported within } [-\tau_p, 0)\cup (0, \tau_p].
\end{align}

Let $\bdelta$ be any testing procedure such that $\mfdrend \le \alpha(1+o(1))$ for any given $\alpha$.
If $\zeta$ satisfies $\vartheta-r<\zeta<\vartheta+r$, then
\[
 p^{-\vartheta-\zeta}\mfdrend + \mfnrend  \ \ge \  L_p p^{-\frac{r^2+(\vartheta-\zeta)^2+2r(\vartheta+\zeta)}{4r}},
\]
where $L_p$ is a multi-log(p) term.
\end{theorem}

The above theorem provides a bound for the weighted sum of the \mfdr and the \mfnrend. In practice, one usually wants to control the \mfdr at a given level $\alpha$, and the following theorem shows that \mfnr could not converge to zero arbitrarily fast.

\begin{theorem}\label{thm:mfnr:rate}
Assume Model (\ref{eqn:model:2}) with conditions \eqref{eqn:LB:unit:diag} and \eqref{eqn:LB:signal}, then for any testing procedure with $\mfdrend\le \alpha$ and for any given $\varkappa>0$, the \mfnr of this procedure satisfies
\[
\mfnrend \ge
 L_p p^{-[\vartheta+ (\sqrt{r}-\sqrt{\vartheta})^2]-\varkappa}.
\]
\end{theorem}
Note the information about $\vX$ has been absorbed. 
This theorem indicates that  $L_p p^{-[\vartheta+ (\sqrt{r}-\sqrt{\vartheta})^2]}$ is essentially the lower bound for \mfnr up to an arbitrarily small penalty.
A testing method is called rate optimal if the \mfnr converges to zero at the rate given in this theorem subject to the condition that the \mfdr is controlled at $\alpha$ level asymptotically.

When $r<\vartheta$, the \mfdr of the decision rule (\ref{eqn:optimal:rule}) converges to one for any $\lambda>0$. We will therefore focus on the case when $r>\vartheta$ for the rest of the paper.
When setting $\lambda=1$, the \mfnr of the oracle decision rule satisfies
\begin{equation*}
\mfnr\approx \frac{\sum_j \theta_j(1-\delta_j)}{p} \ge L_pp^{-\frac{(r+\vartheta)^2}{4r}}.
\end{equation*}
It is easily shown that
\[
\vartheta+(\sqrt{r}-\sqrt{\vartheta})^2 > \frac{(r+\vartheta)^2}{4r}.
\]
This means that the UPS is not rate optimal. The reason is that UPS tries to minimize the Hamming error which assume equal penalty on the type I and type II errors. The resultant \mfdr converges to zero.
Allowing the type I errors by assigning less weight on false discoveries can improve the convergence rate of the \mfnr and increase the power, which is beneficial especially in models with rare signals.
In the next section, we will show that the rate given in Theorem \ref{thm:mfnr:rate} is sharp.


\section{The UPT Method and Upper Bound}\label{sec:upper}
Previously, we have derived the lower bound of \mfnr under Model (\ref{eqn:model:2}) for any procedure which can control \mfdr at a given level.
In this section, motivated by the UPS procedure for variable selection, we will provide a UPT procedure, short for Univariate Penalization for Testing.
Under certain conditions, this method can control \mfdr at $\alpha$ level asymptotically and achieve the optimal convergence rate in \mfnrend.

Let $\hat{\Omega}^*=\vX^T\vX$ and $\hat{\Omega}$ be a matrix such that $\hat{\Omega}_{i,j}=\hat{\Omega}^*_{i,j} \cdot 1\{ |\hat{\Omega}^*_{i,j}|>\log^{-2} (p) \}$.
View $\hat{\Omega}$ as a graph with $p$ node, corresponding to $p$ predictors. Two nodes $i$ and $j$ are connected if and only $\hat{\Omega}_{i,j}\neq 0$.
The UPT method can be summarized as the following steps.
\begin{enumerate}
\item Calculating the marginal correlation $(\vX_{\cdot j}, \vY)$ and keep those predictor with $|(\vX_{\cdot j},\vy)|>t_1$ where $t_1=\sqrt{2q\log{p}}$. Denote this set as $\mathcal{U}_p$, also viewed as a graph induced from $\hat{\Omega}$;
\item Decompose $\mathcal{U}_p$ into small components; 
\item For each component 
 $\mathcal{I}_0\vartriangleleft \mathcal{U}_p$, find the estimator for $\beta_{\mathcal{I}_0}$ to maximize
  \begin{equation}\label{eqn:def:romp}
    \frac{1}{2}\left[ (\vX'\vY)^{\mathcal{I}_0}-(\vX'\vX)^{\mathcal{I}_0}\beta\right]'\left( (\vX'\vX)^{\mathcal{I}_0,\mathcal{I}_0}\right)^{-1} \left[ (\vX'\vY)^{\mathcal{I}_0}-(\vX'\vX)^{\mathcal{I}_0}\beta\right] + t_2^2||\beta||_{\mathcal{I}_0},
  \end{equation}
  subject to a constraint that each coordinate in $\beta_{\mathcal{I}_0}$ is either 0 or $t_3$.
\end{enumerate}

The steps of the UPT method is the same as the UPS method with the key and essential difference lying in the choice of $(t_1,t_2, t_3)$.

For the UPS, these tuning parameters are chosen to balance the type I error and type II error with equal weights and both the type I and type II errors converge to zero.
When considering the testing, we want to maximize the power subject to controlling the type I error at a constant level $\alpha$. This requires much more delicate choices of the tuning parameters.
In multiple testing, the most important, yet the most challenging task is to choose appropriate cutoff value.
For instance, after obtaining the p-values, how to set an cutoff for the rejection region to maximize the power subject to controlling the type I error becomes the key in various multiple testing procedures.
This task is extremely challenging for the high dimensional regression models and none of the existing methods, including UPS, can work well.
The UPT method provides a way to overcome the obstacle which differentiates it from all the existing methods.
We will give details in the rest of this section on how to choose these parameters and theoretically show its validity.

We consider the same conditions about the model as in \cite{Ji:Jin:2012} and restate these conditions in order to better present our result.
Assume that
\begin{align}
\label{eqn:random:design}
\vX_{i\cdot}^T\iid N(0, \frac{1}{n}\Omega)
\end{align}
The conditions for $\Omega$ are summarized as following.
Fixing $A>0$ and $\gamma\in (0,1)$, let
\begin{equation*}
\mathcal{M}^*_p(\gamma, A)= \{\Omega: \sum_{j=1}^p |\Omega(i,j)|^\gamma \le A, \forall 1\le i\le p\}.
\end{equation*}
For any $\Omega\in \mathcal{M}^*_p$, let $U$ be the upper triangular part of $\Omega$ not including the diagonals, and $d(\Omega) = \max\{ ||U(\Omega)||_1, ||U(\Omega)||_\infty\}$.
Fixing $\omega_0\in(0,1/2)$, we consider the following set of correlation matrix
\begin{equation}\label{eqn:def:mp}
\mathcal{M}_p=\{\Omega\in \mathcal{M}^*_p(\gamma, A): d(\Omega)\le \omega_0\}.
\end{equation}
In our study, we also assume that
\begin{align}
n=n_p=p^{\varphi }, \mbox{ with } 1-\vartheta<  \varphi <1,
\end{align}
which is almost necessary for successful variable selection \citep{Donoho:2006}.
Suppose that the support of signal distribution $h_1$ is contained in
\begin{equation}\label{eqn:cond:taup}
[\tau_p, (1+\eta)\tau_p],
\end{equation}
where $\eta$ is defined as
\begin{equation*}
\eta= \frac{\vartheta r}{(\vartheta+r)\sqrt{1+2\omega_0}}\min\{ \frac{2\vartheta}{r}, 1-\frac{\vartheta}{r}, \sqrt{2(1-\omega_0)}-1+\frac{\vartheta}{r}\}.
\end{equation*}

We start with the case where the parameters $(r, \vartheta)$ are known.

\begin{theorem}\label{mfdr:upper:bound}
Assume Model (\ref{eqn:model:2}) and conditions \eqref{eqn:random:design}--\eqref{eqn:cond:taup}. For some fixed $\zeta>0$, set $t_1=\sqrt{2q\log(p)}$ , $t_2=\sqrt{2(\vartheta-\zeta)\log(p)}$, $t_3=\sqrt{2r\log(p)}$, where $\vartheta$ and $q$ are chosen such that $(r+\vartheta-\zeta)^2\ge 4\vartheta r$ and $0<q\le \frac{(r+\vartheta-\zeta)^2}{4r}$. Then the \mfdr of the UPT method satisfies
\[
\mfdr \leq L_pp^{-\frac{r^2+(\vartheta-\zeta)^2-2r(\vartheta+\zeta)}{4r}},
\]
and the \mfnr  satisfies
\[
\mfnr \leq L_pp^{-\frac{r^2+(\vartheta-\zeta)^2+2r(\vartheta+\zeta)}{4r}}.
\]
\end{theorem}
The parameter $t_3$, relating to the signal strength, is the same as that in the UPS. The parameters $t_1$ and $t_2$ depends on $\lambda$ via $\zeta$.
For any $0\le \psi<r-\vartheta$, one can choose $\zeta=r+\vartheta-2\sqrt{r(\vartheta+\psi)}$ so that \mfdr converges to zero in the order of $p^{-\psi}$.
Especially, if we choose $\zeta=(\sqrt{r} -\sqrt{\vartheta} )^2$, the \mfdr is controlled by a single $L_p$ term.
However, this is not enough because in 
testing, 
one usually wants to control \mfdr at $\alpha$ level, asymptotically. Namely, we should choose $t_2$ appropriately such that the multi-log(p) term in Theorem \ref{mfdr:upper:bound} becomes $\alpha(1+o(1))$.



\begin{theorem}\label{thm:upper:mfdr}
 Assume Model (\ref{eqn:model:2}) and conditions \eqref{eqn:random:design}--\eqref{eqn:cond:taup}. For any $\alpha>0$, let  $t_1$ and $t_3$ be the same as in Theorem \ref{mfdr:upper:bound} and
\begin{equation}\label{mfdr:choose:lambda}
t_2^*=\sqrt{ 2(\vartheta - \zeta)\log{p} +  \frac{4r}{r+\vartheta -\zeta} \cdot \left[ (K-\frac{1}{2}) \log\log p - \log M \right] },
\end{equation}
where $\zeta=(\sqrt{r}-\sqrt{\vartheta})^2$, $K$ is a sufficiently large constant,
\[
M=  \frac{\alpha \sqrt{\pi} (r+\vartheta-\zeta) }{ (2e)^K \sqrt{r} (1-\alpha)}  .
\] Then
\[
\mfdr \le \alpha( 1+o(1)) \ \mbox{ and } \ \mfnr \le L_p p^{-\vartheta + (\sqrt{r}-\sqrt{\vartheta})^2}.
\]
\end{theorem}

In the UPS, the tuning parameter $t_2$ only include the $\log(p)$ term and the resultant \mfdr level converges to zero.
The technical proofs in there incorporate some elements from graph theory and are nonconventional.
Here in order to control the \mfdr at a constant level, we must include higher order term $\log{\log{p}}$ and even the constant.
Very different from those for the UPS, the technical proofs here involve sharper asymptotics and
is interesting in its own right.


Similar to the UPS method, the key parameters $(t_1, t_2, t_3)$ can be estimated using the data. Let $\tilde{\vY}=\vX^T\vY$.
Denote the largest off-diagonal coordinate of $\Omega$ by $\delta_0=\delta_0(\Omega)=\max_{\{1\le i,j\le p, i\neq j\}}|\Omega(i,j)|$.
Fix $q$ such that $\max\{\delta_0^2(1+\eta)^2 r, \vartheta-\zeta \} <q \le \frac{ (r+\vartheta-\zeta)^2}{4r}$ and $t_1=\sqrt{2q\log{p}}$.
Define two estimates of $\vartheta$ and $r$ as
\begin{eqnarray}\label{eqn:tuning:estimate}
  \left\{\begin{array}{c} \hat{\vartheta} = \frac{ -2 log( \bar{F}_p( t_1 )) }{2\log{p}},\\
      \hat{r}=\frac{ ( \mu_p(t_1)/\bar{F}_p( t_1 ) )^2}{2\log{p}},
    \end{array}\right.
\end{eqnarray}
where $\mu_p(t)=\frac{1}{p}\sum_j\tilde{Y}_j\cdot 1(\tilde{Y}_j>t)$ and $\bar{F}_p(t)=\frac{1}{p}\sum_j1(\tilde{Y}_j>t)$.


\begin{theorem}\label{thm:upper:mfdr:data}
Assume the same condition in Theorem \ref{thm:upper:mfdr} and the mean of $h(\beta)_j\le \tau_p(1+o(1))$.  Fix $q$ such that $\max\{\delta_0^2(1+\eta)^2 r, \vartheta-(\sqrt{r}-\sqrt{\vartheta})^2 \} <q \le \vartheta$ and $t_1=\sqrt{2q\log{p}}$. Estimate $t_2^*$ in (\ref{mfdr:choose:lambda}) by $\hat{t}_2^*$ using $\hat\vartheta$ and $\hat r$ from (\ref{eqn:tuning:estimate}), and let $ \hat t_3= \sqrt{2 \hat r \log p}.$
Then
\[
\mfdr \le \alpha( 1+o(1))
\quad \mbox{   and   } \quad \mfnr \le L_p p^{-\vartheta + (\sqrt{r}-\sqrt{\vartheta})^2}.
\]
\end{theorem}
In other words, the UPT method achieves the optimal convergence rate in \mfnrend, subject to the controlling of \mfdr at any designated level $\alpha$.

The UPT method depends on two parameters $K$ and $q$. Choosing a sufficiently large $K$ can guarantee the controlling of \mfdr and will not reduce the convergence rate of \mfnrend. We recommend using the maximal component size after thresholding the Gram matrix $\vX^T\vX$.  Similar to the UPS \citep{Ji:Jin:2012}, both theory and simulation studies show the procedure allows some flexibility in the choice of $q$. Running the screening step or even iterating the entire procedure a few times is recommended for best performance.

\section{Simulation}\label{sec:simulation}

We have conducted a few numerical experiments to compare the performance of the UPT method, the BH method and the BY method \citep{Benjamini:Yekutieli:2001} 
for some configurations of $(\vartheta, \theta, \pi_p, \Omega)$. The \mfdr level we tried to control is 0.05. The experiments contain the following steps:
\begin{enumerate}
\item[(1)] Generate a $p\times 1$  vector $\beta$  by   $\beta_j \stackrel{iid}{\sim} (1 - \eps_p) \nu_0 + \eps_p \pi_p$. The distribution $\pi_p$ is taken as a uniform distribution centered at $\tau_p$, and then is assigned a random sign.
\item[(2)] Generate an $n_p\times p$ matrix $X$, the rows of which are  samples  from $N(0, \frac{1}{n_p} \Omega)$;  generate a $n_p\times 1$ vector  $z \sim N(0, I_{n_p})$; let  $Y = X \beta + z$.
\item[(3)] Apply the UPT method and the BH method and the BY method. First, we run the UPT method with the ideal tuning parameters (UPT* method) and the estimated tuning parameters (UPT method).
We choose $K=5$ if it is needed. Second, we fit a simple linear regression between $Y$ and $X_j$ and obtain the usual test statistic $t_j$ and P-value $p_j$ and apply the BH method and the BY method respectively.
\item[(4)] Repeat (2)--(3)  for $100$ independent cycles,  and calculate the average number of true positives (\atpend), the average number of false positives (\afpend) and the false discovery rate (\fdrend).
\end{enumerate}

Define the covariance matrix $\Omega$ as a block-diagonal matrix as
\begin{equation}\label{def:omega}
  \Omega=\left( \begin{array}{cccc} D & 0&\cdots & 0\\
      0&D&\cdots&0\\
      \vdots& \vdots&\vdots&\vdots\\
      0&\cdots&0&D\\
    \end{array}\right), \textrm{where } D=\left( \begin{array}{cc} 1 & a \\a & 1\end{array}\right).
\end{equation}

{\it Experiment 1.} In this experiment, we choose $p=5000$ and $n=1000$, and $\Omega$ as the block diagonal matrix 
$
I_{p/2}\bigotimes \left( \begin{array}{cc} 1 & a \\a & 1\end{array}\right)
$
with $a=0.5$ where $I_{p/2}$ is a $p/2$-dim unit matrix. Let $\vartheta = .5$, and $\pi_p$ as the point mass at $\tau$ which vary from  2 to 8. There are approximately 70 signals, and each is given a random sign. The \atpend, \afpend, and \fdr for each procedure are listed in Table \ref{tab:simul:block}.
\begin{table}

\caption{\label{tab:simul:block}Experiment 1: Comparison for the Block Diagonal $\Omega$ }
\begin{tabular}{c c c c c c}
\hline
& $\tau_p$  &2 & 4 & 6 & 8\\
 \hline
& &  \atp \afp \mfdr & \atp \afp \mfdr & \atp \afp \mfdr & \atp \afp \mfdr \\
BH & &  0.55 0.05 0.08  &  11.20 0.95 0.08 &  23.20 2.85 0.11 & 32.45 5.25 0.14  \\
BY &&  0.10 0.00 0.00 &  \mbox{\ }3.00 0.05 0.02 &   \mbox{\ }7.60 0.20 0.03 &  12.95 0.35 0.03 \\
UPT*&& 1.17 0.10 0.08 & 16.54 1.02 0.06 & 27.98 1.43  0.05 & 37.51 2.58 0.06   \\
UPT && 1.15  0.09  0.08 & 14.97 1.04  0.06  & 26.34 1.56  0.06 & 34.32 2.64  0.06  \\
\hline
\end{tabular}

\end{table}

{\it Experiment 2. } We keep all the parameters as in Experiment  1 but add some random perturbations from Uniform$[-0.5, 0.5]$ to the signals,
and take $\Omega$  as  the penta-diagonal matrix $\Omega(i,j) = 1\{ i = j \}  + 0.5 \cdot  1\{ |i - j| = 1\}  + 0.1 \cdot 1\{ |i - j| = 2\}$.
The results for each procedure are shown in Table \ref{tab:simul:penta}.

\begin{table}

  \caption{\label{tab:simul:penta}Experiment 2: Comparison for the Penta-diagonal $\Omega$.}
  \begin{tabular}{c c c c c c}
    \hline
    & $\tau_p$  &2 & 4 & 6 & 8\\
    \hline
    & &  \atp \afp \mfdr & \atp \afp \mfdr & \atp \afp \mfdr & \atp \afp \mfdr \\
    BH& &   0.50  0.05 0.09 &  10.90  1.55 0.12 &  23.75  4.05 0.15 &  31.95  8.00 0.20  \\
    BY & &  0.10 0.00 0.00 &  \mbox{\ }  3.60 0.25 0.06 &   \mbox{\ }6.50 0.20 0.03 &  12.55 1.05 0.08 \\
    UPT*&& 1.15 0.08 0.07 & 14.95 0.91  0.06 & 26.85  1.46 0.05 & 36.89  2.09 0.06 \\
    UPT && 1.17 0.09  0.07 &  13.87  0.84  0.06 & 26.16 1.51  0.06& 33.72  2.59  0.07\\
    \hline
  \end{tabular}

\end{table}

{\it Experiment 3.} We keep all the parameters as in Experiment 2, but consider the UPT method only with a few choices for the tuning parameter $t_1$, indicated by a factor from 1.10 to 0.90. The results in Table \ref{tab:simul:sensitivity} show that the UPT* method with non-ideal tuning parameters can still outperform the BH and BY methods in Experiment 2, and the procedure itself is not very sensitive to the choice of this threshold. Therefore, in practice, we may try a few values for this threshold or do some kind of iteration.

\begin{table}
  \caption{\label{tab:simul:sensitivity}Experiment 3: Results of the UPT* method with Different Tuning Parameters.}
  \centering
  \begin{tabular}{c c c c c c}
    \hline
    & $\tau_p$  &2 & 4 & 6 & 8\\
    \hline
    & &  \atp \afp \mfdr & \atp \afp \mfdr & \atp \afp \mfdr & \atp \afp \mfdr \\
    1.10 && 0.57 0.05 0.08 & 12.31 0.82 0.06 & 23.48  1.31 0.06 & 31.40  1.47 0.05\\
    1.05 && 0.92 0.07 0.07 & 13.26 0.87  0.06 & 25.69  1.41 0.06 & 34.49  1.86 0.06\\
    1.00 && 1.15 0.08 0.07 & 14.95 0.91  0.06 & 26.85  1.46 0.05 & 36.89  2.09 0.06 \\
    0.95 && 1.18 0.09 0.07 & 15.36 0.98  0.06 & 27.81  1.72  0.07& 37.12  2.56 0.07\\
    0.90 && 1.20 0.09 0.07 & 15.96 1.15 0.07 & 29.56  2.77  0.07& 40.87  2.81 0.06\\
    \hline
  \end{tabular}

\end{table}

{\it Experiment 4.} We keep all the parameters as in Experiment 2. We use a non-Gaussian design for $X$.  In detail,  we generate an $n \times p$ matrix
$M$, the coordinates  of which are iid samples from $\mathrm{Uniform}(-\sqrt{3}, \sqrt{3})$. Second,  generate $\Omega$ as in Experiment 2. Last, let $X = (1/\sqrt{n}) M   \Omega^{1/2}$. The results in \ref{tab:simul:nonGaussian} suggest that the procedure works for more general designs.

\begin{table}

  \caption{\label{tab:simul:nonGaussian}Experiment 4: Results for Non-Gaussian Design }
  \begin{tabular}{cc c c c c}
    \hline
    & $\tau_p$  &2 & 4 & 6 & 8\\
    \hline
    & &  \atp \afp \mfdr & \atp \afp \mfdr & \atp \afp \mfdr & \atp \afp \mfdr \\
    BH&   & 0.54 0.04  0.07 &10.84 1.15 0.10 & 24.54 3.95 0.14 &   31.48 7.64 0.20\\
    BY &  & 0.09 0.06  0.04 &3.52 0.27 0.07 &  7.35 0.44 0.06& 13.58 1.42 0.09     \\
    UPT*&& 1.17 0.09 0.07  &15.63  1.09 0.07&26.81 1.53 0.05 &37.15 2.19  0.06 \\
    UPT && 1.12 0.08 0.07  &14.84 0.97 0.06&26.45   1.46 0.05& 35.12   2.37 0.06\\
    \hline
  \end{tabular}

\end{table}

In Experiments 1, 2, and 4, the proposed UPT method, controls the \mfdr at the $\alpha$-level well; however, its competitor, the BH method, fails to do so. The \mfdr of BH can be as large as 20\%.
The UPT method generally has a larger \atp and a smaller \afp than BH method.
For instance, when $\tau_p=8$ in Experiment 4, the \atp of the UPT method is 4 more than that of BH method. This number is significant given that the signals are rare and weak.
In all the settings, the BY method is too conservative in terms of controlling the \mfdr and the \atp is too low.

In these three experiments, the UPT method has smaller \mfdr than that of the BH method. We expect to discover even more true positives if we set the \mfdr the same.
This is done in the next experiment where we keep the setting in Experiment 4 but adjust the nominal \mfdr level of the UPT method as the empirical \mfdr that of the BH method.
The difference of \atp for these two methods becomes more significant.

{\it Experiment 5.} We keep all the settings in Experiment 4 but adjust the nominal \mfdr level of the UPT* and UPT such that \mfdr is the same as that of the BH method. The ATP and AFP are shown in Table \ref{tab:simul:nonGaussianSamemFDR}. It is shown that both the UPT* and the UPT
discover significantly more signals than the BH method.

\begin{table}

  \caption{\label{tab:simul:nonGaussianSamemFDR}Experiment 5: Results for Non-Gaussian Design for the Same Observed \mfdr}
  \begin{tabular}{cc c c c c}
    \hline
    & $\tau_p$  &2 & 4 & 6 & 8\\
    \hline
    & &  \atp \afp \mfdr & \atp \afp \mfdr & \atp \afp \mfdr & \atp \afp \mfdr \\
    BH&   & 0.54 0.04  0.07 &10.84 1.15 0.10 & 24.54 3.95 0.14 &   31.48 \mbox{\ }7.64 0.20\\
    UPT*&& 1.17 0.09 0.07  &16.38  1.82 0.10&32.31 5.26 0.14 &47.24 11.81 0.20 \\
    UPT && 1.12 0.08 0.07  &15.76 1.75 0.10& 30.77  5.01 0.14&41.40 10.35 0.20\\
    \hline
  \end{tabular}

\end{table}

In summary, the UPT method controls the \mfdr well and is more powerful in identifying the true positives. We therefore strongly recommend it for testing the hypotheses in the high-dimensional regression models.

\section{Conclusion}
Doing statistical inference in the high dimensional data analysis has gained much attention. There are many attempts in recent years. However, how to control the overall error rate at a given level is very challenging under the high dimension, not even to mention the maximization of the power.
In this paper, we borrowed ideas from variable selection to do the multiple testing.
Particularly, we consider the high dimensional regression model with rare and weak signals where the usual criteria such as {\it selection consistency} is not possible.
We thus introduce the error rate, such as \mfdr and \mfnrend, in to the variable selection. Unlike the traditional variable selection criteria, the type I error is allowed as long as the type I error rate is controlled at a certain given level.

Under the rare and weak signals regime, we firstly studied the lower bound of the convergence rate of \mfnr for any testing method which controls the \mfdr at a constant level. We then introduce a method, UPT, a generalization of UPS, and proved that this method achieves the optimal convergence rate of \mfnr subject to controlling \mfdr at $\alpha$ level asymptotically. The proposed method is called rate optimal. Because sharper asymptotics are needed to control the error rates, the UPT involves a set of new theoretical results and the technical arguments are much more challenging and interesting in their own rights. Extensive simulations have shown that the proposed UPT method performs better than many well known existing methods.

 \section{Acknowledgements}
The authors are grateful for helpful comments and discussions with Dr. Jiashun Jin.

\bibliographystyle{rss}
\bibliography{zhaozhg}

\section{Appendix}\label{sec:appendix}
Throughout the Appendix, $L_p$ denotes a multi-log(p) term that may change with each appearance.

\noindent {\bf Proof of Theorem \ref{thm:risk}:}
\begin{eqnarray*}
&& EL(\btheta, \bdelta) |\vy  \\
&=& \sum_j \left\{ \lambda \delta_jfdr_j(\vy) + (1-\delta_j) (1-fdr_j(\vy) ) \right\}\\
&=& \sum_j \left\{ (1-fdr_j(\vy))+\delta_j((\lambda+1)fdr_j(\vy)-1) \right\}\\
&=& \sum_j (1-fdr_j(\vy)) + \sum_j 1_{\{ fdr_j(\vy)\le \frac{1}{\lambda+1}\}}\left( (\lambda+1)fdr_j(\vy)-1\right).
\end{eqnarray*}
Take the expectation on both sides.
Since $E(1-fdr_j(\vy))=\pi_1$,
\begin{eqnarray*}
Risk =  \sum_j E\left\{ 1_{\{fdr_j(\vy)\le \frac{1}{1+\lambda}\}}( (\lambda+1)fdr_j(\vy)   - 1 ) +\pi_1\right \}
\end{eqnarray*}
\qed

To prove Theorem  \ref{thm:fdr:fnr}, we need the following lemma.

\begin{lemma}\label{thm:proposition:1}
Assume Model (\ref{eqn:model:2}) with conditions \eqref{eqn:LB:unit:diag} and \eqref{eqn:LB:signal}.
Then, for any decision rule $\bdelta$, 
\begin{equation}\label{eqn:risk:lower}
Risk(\bdelta^{}) \ge \sum_j \left[ \lambda \pi_0 \bar{\Phi}( \lambdapp ) +  \pi_1 \Phi( \lambdapn ) \right].
\end{equation}
Let $\lambda= p^{-\zeta}$, where $\vartheta -r <\zeta < \vartheta +r $.
Then
\begin{equation}\label{eqn:risk:lower:rate}
\frac{Risk(\bdelta^{})}{p}\ge L_pp^{ -\frac{r^2+(\vartheta-\zeta)^2+2r(\vartheta+\zeta)}{4r}},
\end{equation}
where $L_p$ is a multi-log(p) term.
\end{lemma}

\noindent {\bf Proof of Lemma \ref{thm:proposition:1}:} Let $f_0^j(\vy)$ and $f_1^j(\vy)$ be the density function of $\vY$ under $\theta_j=0$ and $\theta_j=1$.
The marginal density of $\vy$ is given as
\begin{equation}\label{eqn:f}
f(\vy) = \pi_0 f_0^j(\vy) + \pi_1 f_1^j(\vy)\quad \textrm{and}\quad fdr_j(\vy)=P(\theta_j=0|\vy)=\frac{\pi_0f_0^j(\vy)}{f(\vy)}.
\end{equation}
Let $\mathcal{R}_j=\{y: fdr_j(\vy)\le \frac{1}{1+\lambda} \}$. Then
\begin{eqnarray}\label{eqn:expectation:fdr}
&&E1_{\{ fdr_j(\vy)\le \frac{1}{1+\lambda}\}} fdr_j(\vy)=\int_{\mathcal{R}_j} fdr_j(\vy)f(\vy)d\vy = \int_{\mathcal{R}_j}\pi_0f_0^j(\vy)d\vy.
\end{eqnarray}
Similarly,
\begin{eqnarray}\label{eqn:expectation:indicator}
&&E1_{\{ fdr_j(\vy)\le \frac{1}{1+\lambda}\}} (1-fdr_j(\vy)) = \int_{\mathcal{R}_j}\pi_1f_1^j(\vy)d\vy.
\end{eqnarray}
Note that $Risk=EL(\btheta,\bdelta)=E(EL(\btheta,\bdelta)|\vy)$ where $EL(\btheta,\bdelta)|\vy$ can be written as
\[
\lambda\delta_jfdr_j(\vy)+(1-fdr_j(\vy))(1-\delta_j).
\]
Then
\[
Risk=\lambda\int_{\mathcal{R}_j}\pi_0 f_0^j(\vy)d\vy + \int_{\mathcal{R}_j^c}\pi_1 f_1^j(\vy)d\vy.
\]
The remaining of the proof is similar to the proof of Theorem 1.1 in \cite{Ji:Jin:2012}.
The difference is that we consider the parameter $\lambda$, which plays a key role in developing the theory regarding the \mfdr and \mfnrend.
We only sketch the steps of the proof here.
Firstly, we can show that
\[
Risk = \sum_j \left[ \frac{\lambda\pi_0+\pi_1}{2} - \frac{1}{2} \int |\lambda\pi_0f_0^j(\vy) - \pi_1f_1^j(\vy)|d\vy. \right ]
\]

Let $\tilde{\bbeta}_j = \bbeta - \beta_je_j$ where $e_j=(0,\cdots, 0,1,0,\cdots, 0)^T$ is a unit vector with all zero entries except the i-th coordinate.
Let $h(y, \tilde{\bbeta}_j, \beta_j)$ be the joint density of $\vY\sim N(\vX (\tilde{\bbeta}_j+\beta_je_j), \vI_n)$. Let $H(\beta_j)$ and $H(\tilde{\bbeta}_j)$ are the cdf of $\beta_j$ and $\tilde{\bbeta}_j$ respectively. Mimic the proof in \cite{Ji:Jin:2012}, we know that
\begin{eqnarray*}
&&\int |\lambda\pi_0f_0^j(\vy) - \pi_1f_1^j(\vy)|d\vy \\
&\le & \int\int\int M(\lambda,\tilde{\bbeta}_j,\beta_j) d\vy dH(\beta_j) dH(\tilde{\bbeta}_j),
\end{eqnarray*}
where $M(\lambda,\tilde{\bbeta}_j,\beta_j)=\int |\lambda \pi_0 h( \vy, \tilde{\bbeta_j}, 0) - \pi_1 h(\vy, \tilde{\beta}_j, \beta_j) |d\vy $. It is easily seen that
$M(\lambda,\tilde{\bbeta}_j, -\beta_j) = M(\lambda,\tilde{\bbeta}_j,\beta_j)$ and this function is increasing with respect to $\beta_j$ for $\beta_j>0$.
Consequently,
\begin{equation}\label{eqn:append:8}
Risk \ge \sum_j \left[ \frac{\lambda\pi_0+\pi_1}{2} - \frac{1}{2} \int M(\lambda,\tilde{\bbeta}_j, \tau_p)dH(\tilde{\bbeta}_j).\right]
\end{equation}
Let $D=\{ y: \pi_1 h(\vy, \tilde{\bbeta}_j, \tau_p) > \lambda \pi_0 h(\vy, \tilde{\bbeta}_j, 0) \} = \{ y: \pi_1 \exp( \tau_p x_j'(\vy-\vx \tilde{\bbeta}_j) -\tau_p^2/2) >\lambda\pi_0\}$.
Then
\begin{eqnarray*}
&&\frac{ \lambda\pi_0+\pi_1 }{2} - \frac{ M(\lambda,\tilde{\bbeta}_j, \tau_p) }{2} = \lambda\pi_0 \int_D h(\vy,\tilde{\bbeta}_j, 0)d\vy + \pi_1\int_{D^c} h(\vy,\tilde{\bbeta}_j, \tau_p )d\vy.
\end{eqnarray*}
Using the same argument as in the proof of Theorem 1.1 in \cite{Ji:Jin:2012}, we know that
\begin{equation}\label{eqn:append:10}
\int_D h(\vy,\tilde{\bbeta}_j, 0)d\vy = \bar{\Phi}( \lambdapp ), \int_{D^c} h(\vy,\tilde{\bbeta}_j, \tau_p)d\vy = \Phi( \lambdapn ).
\end{equation}
Combining (\ref{eqn:append:8}) and (\ref{eqn:append:10}), we can establish (\ref{eqn:risk:lower}).
Next, we will prove (\ref{eqn:risk:lower:rate}). Note that $\lambda = p^{-\zeta}, \pi_1=p^{-\vartheta}$,
\[
\lambdapp = \frac{\sqrt{2}(r+(\theta-\zeta))}{2\sqrt{r}}\sqrt{\log{p}} + \log L_p,
\]
and
\[
\lambdapn = - \frac{\sqrt{2}(r-(\theta-\zeta))}{2\sqrt{r}}\sqrt{\log{p}} + \log L_p,
\]
Note that $r>\theta-\zeta$. According to Mills' ratio
\begin{eqnarray*}
&&\lambda \pi_0\bar{\Phi}(\lambdapp) + \pi_1 \Phi(\lambdapn)\\
&=& L_p p^{-\zeta} \pi_0\phi(\lambdapp) + L_p \pi_1\phi(( \lambdapn) )\\
&=& L_p p^{-\zeta} p^{-\frac{ (r+(\vartheta-\zeta))^2}{4r}} + L_p p^{-\vartheta}p^{-\frac{ (r-(\theta-\zeta))^2}{4r}}=L_p p^{- \frac{ r^2+(\vartheta-\zeta)^2+2r(\vartheta+\zeta)}{4r} },
\end{eqnarray*}
because
\[
\zeta+ \frac{ (r+(\vartheta-\zeta))^2}{4r} = \vartheta + \frac{ (r-(\vartheta-\zeta))^2}{4r} =  \frac{ r^2+(\vartheta-\zeta)^2+2r(\vartheta+\zeta)}{4r} .
\]
\qed



\noindent {\bf Proof of Theorem \ref{thm:fdr:fnr}}\\
Without loss of generality, we only need to consider
the following three cases:
\begin{enumerate}
 \item[(1)] there exist two multi-log terms  $C_1$ and $C_2$ such that
\[
C_1\le \frac{E\sum_j\delta_j}{p^{1-\vartheta}}\le C_2;
\]

\item[(2)] there exists $K_1>0$ such that
$ \frac{E\sum_j\delta_j}{p^{1-\vartheta}}  =o(p^{-K_1}) ; $

\item[(3)] there exists $K_2>0$ such that
$
  \frac{E\sum_j\delta_j}{p^{1-\vartheta}}  \gtrapprox  p^{K_2}  .
$

\end{enumerate}
If case (3) holds, then
\[
\mfdrend = \frac{E\sum_j (1-\theta_j)\delta_j}{E\sum_j\delta_j} = 1+o(1).
\]
This contradicts the assumption that $\mfdrend\le \alpha(1+o(1))$. Consequently, we only focus on case (1) and (2).

In case (1), by Lemma \ref{thm:proposition:1},
\begin{eqnarray*}
&&p^{-\vartheta-\zeta}\mfdr + \mfnr \\
&=& p^{-\vartheta-\zeta} \frac{E\sum_j (1-\theta_j)\delta_j}{E\sum_j\delta_j} + \frac{E\sum_j \theta_j(1-\delta_j)}{E\sum_j(1-\delta_j)} \\
&\gtrapprox & \frac{1}{p}( L_p p^{-\zeta} E\sum_j (1-\theta_j)\delta_j + E\sum_j \theta_j(1-\delta_j) ) \\
&\ge & L_p p^{-\frac{r^2+(\vartheta-\zeta)^2+2r(\vartheta+\zeta)}{4r}} .
\end{eqnarray*}

In case (2),
\[
\mfnrend = \frac{ E\sum_j \theta_j(1-\delta_j)}{E\sum_j (1-\delta_j) } \gtrapprox \frac{ p^{1-\vartheta}}{p} = p^{-\vartheta}.
\]
Then
\[
p^{\frac{r^2+(\vartheta-\zeta)^2+2r(\vartheta+\zeta)}{4r}}( p^{-\vartheta-\zeta}\mfdr + \mfnr) \ge p^{\frac{(r+\zeta-\vartheta)^2}{4r}}\ge L_p.
\]

This completes the proof. \qed


\noindent {\bf Proof of Theorem \ref{thm:mfnr:rate}:}
\\
For any sufficiently small $\varkappa>0$, let
\[
\varkappa'=\vartheta - \left[\sqrt{r}-\sqrt{ (\sqrt{r}-\sqrt{\vartheta})^2 +\varkappa} \right]^2>0.
\]
Consequently,
\[
(\sqrt{r}-\sqrt{\vartheta-\varkappa'})^2 = (\sqrt{r}-\sqrt{\vartheta})^2 + \varkappa.
\]
Let $\zeta= r+\vartheta-2\sqrt{r(\vartheta-\varkappa')}$.
As a result,
\begin{eqnarray*}
&& \frac{r^2+(\vartheta-\zeta)^2 + 2r(\vartheta+\zeta)}{4r} - \vartheta -\zeta \\
&=&\frac{(r+\vartheta-\zeta)^2-4r\vartheta}{4r} = \frac{4r(\vartheta-\varkappa')-4r\vartheta}{4r}\\
&=&-\varkappa'.
\end{eqnarray*}
This implies that
\[
p^{\frac{r^2+(\vartheta-\zeta)^2 + 2r(\vartheta+\zeta)}{4r} - \vartheta -\zeta }\mfdrend = O(p^{-\varkappa'}).
\]
According to Theorem \ref{thm:fdr:fnr},
\begin{equation}\label{app:eqn:mfnr}
p^{\frac{r^2+(\vartheta-\zeta)^2+2r(\vartheta+\zeta)}{4r}} \mfnrend \ge L_p.
\end{equation}
Note that
\begin{eqnarray*}
&& \frac{r^2+(\vartheta-\zeta)^2+2r(\vartheta+\zeta)}{4r} \\
&=&\frac{ (\zeta-\vartheta+r)^2+4r\vartheta}{4r} = \frac{ \left[ 2\sqrt{r}(\sqrt{r}-\sqrt{\vartheta-\varkappa'})\right]^2 + 4r\vartheta } {4r} \\
&=&(\sqrt{r}-\sqrt{\vartheta-\varkappa'})^2+\vartheta = \vartheta+(\sqrt{r}-\sqrt{\vartheta})^2 + \varkappa.
\end{eqnarray*}
Combining this with equation (\ref{app:eqn:mfnr}), we know that
\[
\mfnrend \ge L_p p^{-\left[ \vartheta+(\sqrt{r}-\sqrt{\vartheta})^2\right] -\varkappa}.
\]
\qed

In order to prove Theorem \ref{mfdr:upper:bound}, we need the following lemmas.
\begin{lemma}\label{prop:screen}
 Assume Model (\ref{eqn:model:2}) and conditions \eqref{eqn:random:design}--\eqref{eqn:cond:taup}. If $0<q\le \frac{(r+\vartheta-\zeta)^2}{4r}$, then
\[
\frac{1}{p}\sum_{j=1}^p P( \vx_{j\cdot}'\vY<t_1, \beta_j\neq 0) \le L_p p^{-\frac{r^2+(\vartheta-\zeta)^2+2r(\vartheta+\zeta)}{4r}} .
\]
\end{lemma}


\begin{lemma} \label{prop:sas}
 Assume Model (\ref{eqn:model:2}) and conditions \eqref{eqn:random:design}--\eqref{eqn:cond:taup}. As $p \rightarrow \infty$, there is a constant $K$ such that with probability $1- o(p^{-\frac{r^2+(\vartheta-\zeta)^2+2r(\vartheta+\zeta)}{4r}})$, each connected component of $\mathcal{U}_p$ has no more than $K$ nodes.
\end{lemma}

\noindent {\bf Proof of Lemma \ref{prop:screen}:}
\\
According to the proof of Lemma 2.1 in \cite{Ji:Jin:2012}, with probability of $1+o(1/p^D)$ where $D$ is a sufficiently large constant,
\begin{eqnarray*}
P(x_j'Y<t, \theta_j=1) \le p^{-\vartheta}\Phi( t-\tau_p).
\end{eqnarray*}
According to Mill's ratio, the right hand side can be simplified as
$ L_p p^{-\vartheta - (\sqrt{r}-\sqrt{q})^2}$ which is smaller than or equal to  $L_p p^{-\frac{r^2+(\vartheta-\zeta)^2+2r(\vartheta+\zeta)}{4r}}$ because $0<q\le \frac{(r+\vartheta-\zeta)^2}{4r}$.
\qed

\begin{lemma}\label{upper:bound}
 Assume Model (\ref{eqn:model:2}) and conditions \eqref{eqn:random:design}--\eqref{eqn:cond:taup}. Set $t_1=\sqrt{2q\log(p)}$, $t_2=\sqrt{2(\vartheta-\zeta)\log(p)}$, $t_3=\sqrt{2r\log(p)}$ for the UPT method, then the weighted classification error associated with the loss function (\ref{eqn:loss}) where $\lambda= L_pp^{-\zeta}$ is given as
\[
\frac{Risk}{p} \le L_p p^{-\frac{r^2+(\vartheta-\zeta)^2+2r(\vartheta+\zeta)}{4r}}.
\]
\end{lemma}

\noindent {\bf Proof of Lemma \ref{upper:bound}.}
Define the event $A_p$ as
\begin{eqnarray*}
A_p&=&\{ |(X'X)(i,j)-\Omega(i,j)|\le L_p p^{-\omega/2}, \forall 1\le i,j\le n\}\\
&\cap& \{ ||(X'X)^{\mathcal{I}_0, \mathcal{I}_0}-\Omega^{\mathcal{I}_0, \mathcal{I}_0}||_\infty \le L_pp^{-\omega/2}\}.
\end{eqnarray*}
According to \cite{Ji:Jin:2012}, $P(A_p)=1-o(1/p^D)$ where $D$ is a sufficiently large constant. Consequently, we only need to show the result when $X\in A_p$.

Now, the risk can be naturally written as two parts $Risk = I +II$ where
\[
I = \sum_j E( L(\theta_j,\delta_j) 1(j \notin \mathcal{U}_p(t))|X),
\quad\quad II = \sum_j E( L(\theta_j,\delta_j) 1(j\in \mathcal{U}_p(t))|X).
\]
According to Lemma \ref{prop:screen},
\[
I = \sum_j P(\theta_j=1, \delta_j=0, j \notin \mathcal{U}_p(t)) \le L_pp\cdot p^{-\optrate}.
\]
We only need to prove that $II\le L_p p \cdot p^{-\optrate}$.

The remaining proof is similar to the proof of Theorem 2.1 in \cite{Ji:Jin:2012}. The difference lies in the extra parameter $\lambda$. We only sketch the main steps here.
According to the proof of Lemma 2.3 of \cite{Ji:Jin:2012}, there exists a constant $K>0$ and event $A_p$ such that $P(A_p^c)\le o(1/p^D)$ and that any subgraph of $\mathcal{U}_P(t)$ has at most $K$ elements over the event $A_p$. We only need to show that $II\le L_p p \cdot p^{-\optrate}$ over the event $A_p$.
Following the similar argument of the proof of Theorem 2.1 in \cite{Ji:Jin:2012}, we only need to prove that
\[
E L(\theta_j,\delta_j)1(j\in \mathcal{I}_0 \vartriangleleft \mathcal{U}_p(t_p^*)\cap A_p\cap B_p) \le L_p p^{-\optrate}.
\]
where $B_p$ is defined through $B_p^c$ as
\[
B_p^c(\mathcal{I}_0)=\{ \textrm{There are indices $j\notin \mathcal{I}_0$ and $k\in \mathcal{I}_0$ such that $\beta_j\neq 0,\Omega^*(j,k)\neq 0$}\}.
\]
Now, we consider the type I error and type II error separately.
\[
I^* = E1(\theta_j=0, \delta_j=1)1(j\in \mathcal{I}_0 \vartriangleleft \mathcal{U}_p(t_p^*)\cap A_p\cap B_p),
\]
and
\[
II^* = E1(\theta_j=1, \delta_j=0)1(j\in \mathcal{I}_0 \vartriangleleft \mathcal{U}_p(t_p^*)\cap A_p\cap B_p),
\]

Let $B_{nn}$ be the number of true negatives, $B_{ns}$ be the number of false positives, $B_{sn}$ be the number of false negatives, and $B_{ss}$ be the number of true positives within $\mathcal{I}_0$. When considering the above type II error, then $B_{sn}+B_{ss}\ge 1$.
Consequently,
\[
II^* \le L_pp^{-\vartheta (B_{sn}+B_{ss})}\bar{\Phi}( F ).
\]
where $F$ is defined as the right hand side of (A.44) of \cite{Ji:Jin:2012}. It can be further similarly shown that when $B_{sn}+B_{ss}\ge 1$,
\[
II^* \le L_pp^{-\vartheta}p^{-\frac{( r+\vartheta-\zeta)^2}{4r} }.
\]

Next consider type I error where
\[
I^* \le p^{-\vartheta(B_{sn}+B_{ss})}\bar{\Phi}(F).
\]
When $B_{sn}+B_{ss}=0$, then it was shown in \cite{Ji:Jin:2012} that
\[
I^*  \le L_p p^{-\frac{ (r+\vartheta-\zeta)^2}{4r}}.
\]
When $B_{sn}+B_{ss}\ge 1$, then
\[
I^* \le L_p p^{-\vartheta}p^{-\frac{ (r-\vartheta+\zeta)^2}{4r}}.
\]
Note that
\[
\frac{ (r+\vartheta-\zeta)^2}{4r} -\vartheta - \frac{ (r-\vartheta+\zeta)^2}{4r} = -\zeta <0.
\]
Therefore,
\[
I^* \le L_p p^{-\frac{ (r+\vartheta-\zeta)^2}{4r}}.
\]
In summary,
\begin{eqnarray*}
&&E L(\theta_j,\delta_j)1(j\in \mathcal{I}_0 \vartriangleleft \mathcal{U}_p(t_p^*)\cap A_p\cap B_p)\\
&\le & p^{-\zeta}  I^* + II^* \le L_p\left[ p^{-\zeta}p^{-\frac{(r+\vartheta-\zeta)^2}{4r}} + p^{-\vartheta} p^{-\frac{(r+\vartheta-\zeta)^2}{4r}} \right] \\
& = & L_p p^{-\optrate}.
\end{eqnarray*}
\qed

\noindent {\bf Proof of Theorem \ref{mfdr:upper:bound}:}
\\
Note that
\begin{eqnarray*}
&&\mfdr= \frac{ \sum_j P(\theta_j=0, \delta_j=1)}{ \sum_j P(\delta_j=1) } \\
&=& \frac{ \sum_j P(\theta_j=0, \delta_j=1) }{  \sum_j \left[ P(\theta_j=0, \delta_j=1) +  P(\theta_j=1, \delta_j=1) \right] }.
\end{eqnarray*}
In the denominator, note that
\[
 P(\theta_j=1, \delta_j=1) = P(\theta_j=1) - P(\theta_j=1, \delta_j=0) = p^{-\vartheta} -  P(\theta_j=1, \delta_j=0).
\]
According to Lemma \ref{upper:bound},
\[
\sum_j P(\theta_j=1, \delta_j=0) \le Risk \le L_p p\cdot p^{-\frac{r^2+(\vartheta-\zeta)^2+2r(\vartheta+\zeta)}{4r}}.
\]
Consequently,
\[
\sum_j P(\theta_j=1, \delta_j=1) \ge p\left[ p^{-\vartheta}- L_p  p^{-\frac{r^2+(\vartheta-\zeta)^2+2r(\vartheta+\zeta)}{4r}} \right] \ge p^{1-\vartheta}(1+o(1)).
\]
Next, we consider the numerator $P(\theta_j=0, \delta_j=1)$. Note that
\[
\lambda \sum_j P(\theta_j=0, \delta_j=1) \le Risk = L_p p\cdot p^{-\frac{r^2+(\vartheta-\zeta)^2+2r(\vartheta+\zeta)}{4r}}.
\]
Consequently,
\[
\sum_j P(\theta_j=0, \delta_j=1) \le L_p p\cdot p^{-\frac{r^2+(\vartheta-\zeta)^2+2r(\vartheta+\zeta)}{4r}+\zeta } = L_p p\cdot p^{ -\frac{( r+\vartheta-\zeta)^2}{4r}}.
\]
Since $\frac{( r+\vartheta-\zeta)^2}{4r} \ge \vartheta$, then
\[
mFDR\le L_p p^{ -\frac{( r+\vartheta-\zeta)^2}{4r} +\vartheta } = L_p p^{-\frac{r^2+(\vartheta-\zeta)^2-2r(\vartheta+\zeta)}{4r}}.
\]
Note that
\[
\mfnr=\frac{ \sum_jP(\theta_j=1,\delta_j= 0) }{ \sum_j P(\delta_j= 0) }  = \frac{ \sum_j P(\theta_j=1, \delta_j= 0) }{ \sum_j\left[ P(\theta_j=1, \delta_j= 0) +  P(\theta_j= 0, \delta_j= 0)\right] }.
\]
Note that
\[
P(\theta_j=0, \delta_j=0)=P(\theta_j=0)-P(\theta_j=0, \delta_j=1) =\pi_0(1+o(1)).
\]
Consider the numerator
\[
\sum_j P(\theta_j=1, \delta_j=0)\le Risk\le L_pp\cdot p^{-\frac{r^2+(\vartheta-\zeta)^2+2r(\vartheta+\zeta)}{4r}}.
\]
Consequently,
\[
\mfnr \le L_p p^{-\frac{r^2+(\vartheta-\zeta)^2+2r(\vartheta+\zeta)}{4r}}.
\]
\qed

\noindent {\bf Proof of Theorem \ref{thm:upper:mfdr}.} \\
According to the definition, we know that
\[
\mfdr = \frac{\sum_j P(\theta_j=0, \delta_j=1)} {\sum_j P(\theta_j=0, \delta_j=1) + \sum_j P(\theta_j=1, \delta_j=1)}.
\]
According to the proof of Theorem \ref{mfdr:upper:bound},
\[
\sum_jP(\theta_j=1,\delta_j=1)\ge p^{1-\vartheta}(1+o(1)).
\]
This leads to
\begin{align}\label{eq:upper:mfdr}
\mfdr \le \frac{\sum_j P(\theta_j=0, \delta_j=1)} {\sum_j P(\theta_j=0, \delta_j=1) + p\cdot p^{-\vartheta}(1+o(1)) }.
\end{align}
Next, we consider the type I error $\sum_j P(\theta_j=0, \delta_j=1)$.
By Lemma \ref{prop:sas}, there is a constant $K>0$ and an event $A_p$ such that $P(A_p^c) \leq o(p^{-(r+\vartheta-\zeta)^2/(4r)})$.
Within $A_p$, similar to the proof of Theorem 2.1 in \cite{Ji:Jin:2012}, we can show that


\begin{align}\label{append:eqn:10}
&P(\theta_j=0, \delta_j=1, A_p) \\
\le & (2e\log{p})^K\max\limits_{\mathcal{I}_0}\{ P(\theta_j=0, \delta_j=1, j\in\mathcal{I}_0 \vartriangleleft \mathcal{U}_p(t_p^*), A_p) \nonumber \},
\end{align}
where in the last equation $\mathcal{I}_0$ runs over all connected subgraphs of size $\ell \leq K$ and containing $j$.
For any $\mathcal{I}_0$ that is a component of  $\mathcal{U}_p(t_1)$, denoted by $\mathcal{I}_0\vartriangleleft  \mathcal{U}_p(t_1)$, there is an $|\mathcal{I}_0| \times 1$ vector $\tilde z \sim N(0, \Omega^{\mathcal{I}_0, \mathcal{I}_0})$ independent of $\beta^{\mathcal{I}_0}$ such that
\[\tilde Y^{\mathcal{I}_0} = \Omega ^{\mathcal{I}_0, \mathcal{I}_0} \beta^{\mathcal{I}_0} +\tilde z + rem,
\quad ||rem||_\infty \leq o(1/\sqrt{\log p})
\]
and eventually,
\[
P(\theta_j=0, \delta_j=1, j\in\mathcal{I}_0 \vartriangleleft \mathcal{U}_p(t_1), A_p) \le p^{-\vartheta (B_{sn}+B_{ss})}\bar{\Phi}(G),
\]
where
\[
G=\frac{1}{2\tau_p\sqrt{\Delta_1'\Omega\Delta_1}}(-dt_2^{*2} + \tau_p^2\Delta_1'\Omega\Delta_1 + 2\tau_p^2 \Delta_1'\Omega\Delta_2 + o(1/\sqrt{2\log{p}})),
\]
where the last term is non-stochastic with a negligible effect.
The difference between this and (A.42) in \cite{Ji:Jin:2012} is that we require the error term to be smaller than $o(\frac{1}{\sqrt{\log{p}}})$ to ensure the controlling of \mfdr.
This is guaranteed by thresholding the sample covariance matrix $\Omega^*$ at $\frac{1}{\log^2{p}}$.

Similar to the proof in Lemma 6.5 of \cite{Ji:Jin:2012},
\begin{align*}
  G
  \ge \frac{2(\vartheta-\zeta) \log p + 2 r \log p + \frac{4r}{r+\vartheta -\zeta} \cdot \left[ (K-\frac{1}{2}) \log\log p - \log M \right]  }  {2\sqrt{2 r \log p}}
\end{align*}
By Mill's ratio,
\begin{align}
  \bar\Phi(G) \leq  \frac{(1+o(1)) \sqrt{2r \log p} \cdot p^{-(r+\vartheta-\zeta)^2/(4r)}   e^{ (\frac{1}{2}-K) \log\log p - \log M  }}{\sqrt{2\pi} (r+\vartheta-\zeta) \log p }
\end{align}
Combining with \eqref{append:eqn:10} gives
\[ P(\theta_j=0, \delta_j=1, A_p)
\leq
(1+o(1)) \frac{\alpha}{1-\alpha} \cdot   p^{-(r+\vartheta-\zeta)^2/(4r)} .
\]
Consequently,
\begin{eqnarray*}
&&P(\theta_j=0, \delta_j=1)= P(\theta_j=0, \delta_j=1, A_p)+P(\theta_j=0, \delta_j=1, A_p^c)\\
&\le& P(\theta_j=0, \delta_j=1, A_p) + o(  p^{-(r+\vartheta-\zeta)^2/(4r) }) \\
&\le& (1+o(1)) \frac{\alpha}{1-\alpha} \cdot   p^{-(r+\vartheta-\zeta)^2/(4r)} .
\end{eqnarray*}
Combining with \eqref{eq:upper:mfdr} and $\zeta = (\sqrt{r} - \sqrt{\vartheta})^2$ gives
\[
\mfdr \leq
 \frac{ (1+o(1)) \frac{\alpha}{1-\alpha} \cdot   p^{-(r+\vartheta-\zeta)^2/(4r)}}{ (1+o(1)) \frac{\alpha}{1-\alpha} \cdot   p^{-(r+\vartheta-\zeta)^2/(4r)} + p^{-\vartheta} (1+o(1))} = \alpha (1+o(1)).
\]

According to Theorem \ref{mfdr:upper:bound} and $\zeta=(\sqrt{r}-\sqrt{\theta})^2$, then
\[
\mfnr = L_p p^{-(\vartheta+(\sqrt{r}-\sqrt{\vartheta})^2)}.
\]

\qed

\noindent{\bf Proof of Theorem \ref{thm:upper:mfdr:data}}

First, we have the following lemma for estimating the tuning parameters
$t_2^*$ and $t_3$.

\begin{lemma} \label{thm:upper:mfdr:data:estimate}
  Under the conditions of Theorem \ref{thm:upper:mfdr:data}, as $p\rightarrow \infty$ with probability $1-o(1/p^D)$ with a constant $D>0$, there is a nonrandom $g_p=o(1/\sqrt{\log p}) $ such that
  \[ | \hat{t}_2^* - t_2^*  |\leq g_p
\quad \mbox{  and  }  \quad
   | \hat{t}_3 - t_3 | \leq g_p
   .\]
\end{lemma}

Then
\[ (1- g_p) t_2^* \leq \hat t_2^* \leq (1+ g_p) t_2^*,
\quad \mbox{ and } \quad
(1- g_p) t_3 \leq \hat t_3 \leq (1+ g_p) t_3.
\]
By a close investigation of the proof of Theorem 2.3, all the arguments for the \mfdr still hold if we replace $t_2^*$ by $(1 \pm g_p) t_2^*$
and $t_3$ by $(1 \pm g_p ) t_3$.  It is also the case for \mfnr except that the generic log term $L_p$ may be slightly different. Therefore, the proofs follow.

\noindent{\bf Proof of Lemma \ref{thm:upper:mfdr:data:estimate}}

The proof is similar to that of Lemma 2.4 in \cite{Ji:Jin:2012}, except that we need to choose
\begin{align*}
\delta_p &  = 1/(\log p)^2,  \\
a_p & = 1+o(1/\sqrt{\log p}),  \\
b_p & = o(1/\log p),
\end{align*}
and we need to control $\tilde Y$ with probability $1+o(1/p^D)$ for sufficiently large $D>0$. Then by similar techniques, we will have
\[ || W-\tilde Y ||_\infty  \leq b_p \sqrt{2\log p}.
\]

Introduce event $A_p = \{\| \tilde{Y} - W\|_{\infty} \leq b_p \sqrt{2 \log(p)} \}$, and
\[ \bar F^{\pm}_p(t) = \frac{1}{p} \sum_{j = 1}^p  1_{\{ W_j  \pm  b_p \sqrt{2 \log p} \geq t\}}.
\]
Comparing $\bar F^{\pm}_p(t)$ with  $\cf_p(t)$,  it is seen that over the event $A_p$,
\[
\cf^{-}_p(t) \leq  \cf_p(t) \leq \cf^{+}_p(t).
\]
The claim follows from the following lemma, which is proved in \cite{Ji:Jin:2012}.

\begin{lemma}  \label{lemma:cfmain}
Under the conditions of Theorem \ref{thm:upper:mfdr:data}, 
there is a constant $c = c(\vartheta, r) > 0$ such that, with probability $1-o(1/p^D)$,
\[
\bigg| \frac{1}{p \eps_p}  \sum_{j =1}^p 1_{\{ W_j \geq t \}}  - 1 \biggr| \leq  L_p p^{-c(\vartheta, r)}.
\]
\end{lemma}

\qed

\end{document}